\newcommand{\be}{\begin{equation}}
\newcommand{\ee}{\end{equation}}

\documentclass[twocolumn,showpacs,superscriptaddress,amsmath,amssymb]{revtex4}
\usepackage{graphicx}
\usepackage{dcolumn}
\usepackage{bm}

\begin{document}
\title{Self-trapped bidirectional waveguides in a saturable photorefractive medium}

\author{M.~Beli\'{c}}
\affiliation{Institute of Physics, P.O.~Box 57, 11001 Belgrade,
Yugoslavia} \affiliation{Institute of Applied Physics,
Westf\"{a}lische Wilhelms-Universit\"{a}t M\"{u}nster, D-48149 M\"{u}nster,
Germany}
\author{Ph.~Jander}
\affiliation{Institute of Applied Physics, Westf\"{a}lische
Wilhelms-Universit\"{a}t M\"{u}nster, D-48149 M\"{u}nster, Germany}
\author{A.~Strini\'c}
\affiliation{Institute of Physics, P.O.~Box 57, 11001 Belgrade,
Yugoslavia} \affiliation{Institute of Applied Physics,
Westf\"{a}lische Wilhelms-Universit\"{a}t M\"{u}nster, D-48149 M\"{u}nster,
Germany}
\author{A.~Desyatnikov}
\author{C.~Denz}
\affiliation{Institute of Applied Physics, Westf\"{a}lische
Wilhelms-Universit\"{a}t M\"{u}nster, D-48149 M\"{u}nster, Germany}

\begin{abstract}
We introduce a time-dependent model for the generation of joint
solitary waveguides by counterpropagating light beams in a
photorefractive crystal. Depending on initial conditions, beams
form stable steady-state structures or display periodic and
irregular temporal dynamics. The steady-state solutions are
non-uniform in the direction of propagation and represent a
general class of self-trapped waveguides, including
counterpropagating spatial vector solitons as a particular case.
\end{abstract}

\maketitle

During the last decade spatial screening solitons \cite{opn02}
have been considered almost exclusively in copropagation geometry.
Recent progress in generating optical solitons consisting of
counterpropagating fields by Cohen et al. \cite{cohen} has renewed
interest in counterpropagating four-wave mixing, extensively
studied in the past \cite{silberberg,polar,haelt}. However, such
geometries in photorefractive (PR) media are prone to
instabilities \cite{honda,saffman,belic} and are often employed
for transverse optical pattern formation \cite{arecchi}. In
particular, \emph{temporal} instabilities were shown to result in
self-oscillation, chaos, and bistability \cite{silberberg,polar}.
It is therefore of importance to investigate the temporal behavior
of counterpropagating self-trapped beams in PR crystals with
finite response time. Furthermore, one may easily envision
interest in a stable self-adjustable bidirectional connection of
two arrays of beams across a PR crystal.

In this communication we derive equations for the propagation of
beams, similar to the bimodal counterpropagating solitons in Kerr
media \cite{haelt}, and collisions of screening PR solitons
propagating in opposite directions \cite{cohen}, together with a
time-relaxation procedure for the space charge field determining
the refractive index modulation in PR crystals. Dynamical effects
are found important for understanding the behavior of
counterpropagating beams. We display numerically the temporal
formation of bright spatial screening vector solitons formed by
counterpropagating beams, and discuss their interactions in (1+1)
spatial dimensions. Beyond soliton solutions, we introduce a more
general class of steady-state induced waveguides. Additionally, a
situation where the interacting beams do not converge to a
stationary structure, but rather alternate between different
states, is reported.

We consider two counterpropagating light beams in a PR crystal, in
the paraxial approximation, under conditions suitable for the
formation of screening solitons. The optical field is given as the
sum of the counterpropagating waves $F\exp(ikz+i\omega
t)+B\exp(-ikz+i\omega t)$, $k$ being the wave vector in the
medium, $F$ and $B$ are slowly varying envelopes of the beams. The
total light intensity $I$ is measured in units of the background
light intensity, also necessary for the generation of solitons.
After averaging in time on the scale of response time $\tau_0$ of
the PR crystal, the total intensity is given by: \be\label{int}
1+I=(1+I_0)\left[1+\varepsilon\left \{m\exp(2ikz)+c.c.\right
\}/2\right],\ee where $I_0=|F|^2+|B|^2$, $m=2FB^*/(1+I_0)$ is the
modulation depth, and $c.c.$ stands for complex conjugation. Here
the parameter $\varepsilon$ measures the degree of \emph{temporal}
coherence of the beams related to the crystal relaxation time: for
$\varepsilon=0$, i.e.\ when the relative phase of the beams varies
much faster than $\tau_0$, the beams are effectively incoherent
(see discussion in \cite{cohen}). In the opposite case
$\varepsilon=1$, the intensity distribution contains an
interference term which is periodically modulated in the direction
of propagation $z$, chosen to be perpendicular to the $c$-axis of
the crystal, which is also the $x$-axis of the coordinate system.
Beams are polarized in the $x$ direction, and the external
electric field $E_e$, necessary for the formation of self-trapped
beams, also points in the $x$ direction. The electric field in the
crystal couples to the electro-optic tensor, giving rise to a
change in the index of refraction of the form \be \Delta
n=-\frac{n_0^3}{2}r_{eff}E_{tot},\ee where $n_0$ is the
unperturbed index, $r_{eff}$ is the effective component of the
electro-optic tensor (in this case $r_{33}$), and $E_{tot}$ is the
$x$-component of the total electric field. It consists of the
external field and the space-charge field $E_{sc}$ generated in
the crystal, $E_ {tot}=E_e+E_{sc}$.

The intensity modulates the space charge field, which we represent
in the normalized form \be\label{schf}
E_{sc}/E_e=E_0+\frac{1}{2}\left[E_1\exp(2ikz)+c.c.\right] \,\ee
where $E_0$ is the homogeneous part of the $x$-component of the
space charge field, and $E_1(x,z)$ is the additional \emph{slowly
varying} part of the space-charge field proportional to
$\varepsilon$, $|\partial_z E_1|\ll 2k|E_1|$. It is $E_0$ that
screens the external field, and $E_1$ is the result of the
interference pattern along the $z$-direction. It vanishes together
with the intensity modulation for incoherent beams, i.e.\ in the
limit $\varepsilon=0$.
\begin{figure}
\includegraphics[width=75mm]{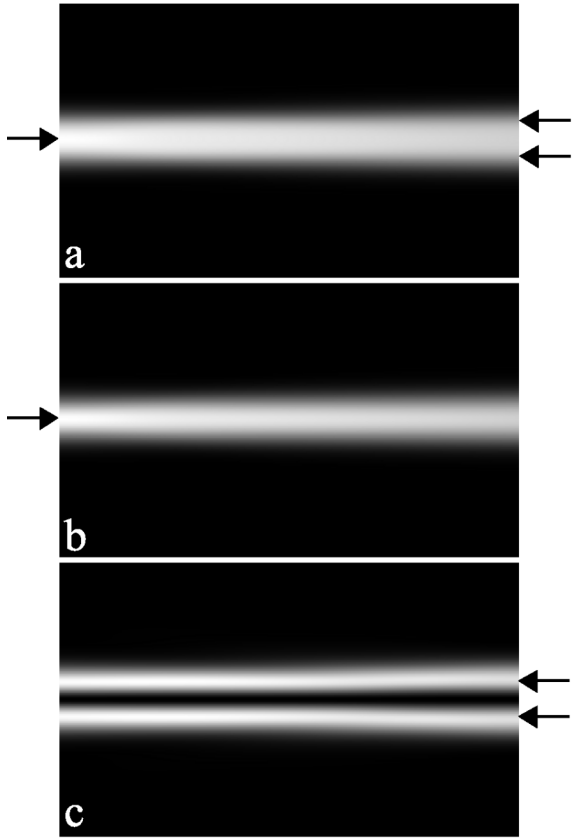}
\vspace{-2mm} \caption{\label{fig1} Counterpropagating dipole-mode
vector soliton (a), made out of a fundamental beam propagating to
the right (b), and a coherent dipole beam propagating to the left
(c). Coupling strength $\Gamma=10/3$.}
\end{figure}
\begin{figure}
\includegraphics[width=75mm]{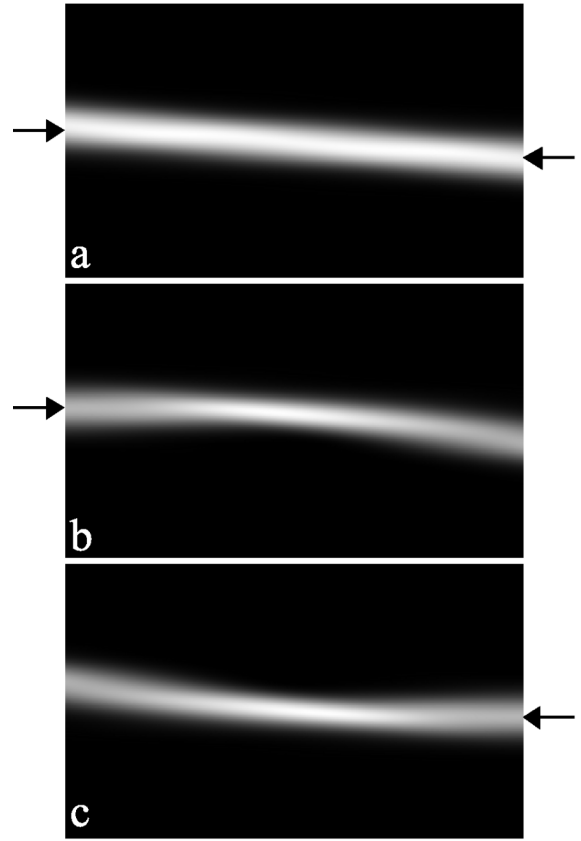}
\vspace{-2mm} \caption{\label{fig2} Bidirectional waveguide. (a)
Total intensity distribution; (b) right- and (c) left-propagating
beams. Layout as in Fig.~\ref{fig1}, parameters $\varepsilon=1$,
and $\Gamma=5$. Initial peak intensities $I_F=I_B=1$.}
\end{figure}

In a simplified approach, one assumes a local, isotropic
approximation to the space charge field, and looks for a solution
with saturable nonlinearity $E_{sc}=E_e/(1+I)$. Substituting
Eqs.~(\ref{int}) and (\ref{schf}) in this expression, neglecting
higher harmonics and terms quadratic in $m$, we obtain as a
steady-state solution \be E_0=-\frac{I_0}{1+I_0}\ ,\qquad
E_1=-\frac{\varepsilon m}{1+I_0}\ .\ee

Temporal evolution of the space charge field is introduced by
assuming relaxation-type dynamics \cite{solymar}
\begin{subequations}\label{temp}\begin{eqnarray}
\tau\partial_tE_0+E_0=-\frac{I_0}{1+I_0}\ ,\\
\tau\partial_tE_1+E_1=-\frac{\varepsilon m}{1+I_0}\
,\end{eqnarray}\end{subequations} where the relaxation time of the
crystal $\tau$ is inversely proportional to the total intensity
$\tau=\tau_0/(1+I)$, i.e.\ illuminated regions in the crystal
react faster. The assumed dynamics is that the space-charge field
builds up towards the steady-state, which depends on the light
distribution, which in turn is slaved to the slow change of the
space-charge field. As it will be seen later, this does not
preclude a more complicated dynamical behavior.
\begin{figure}
\includegraphics[width=85mm]{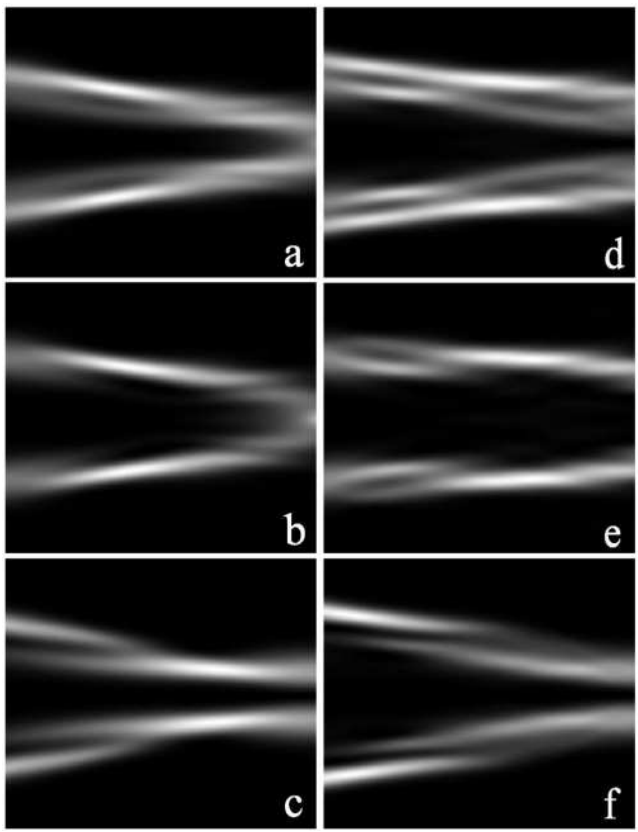}\vspace{-2mm}
\caption{\label{fig3} Incoherent (a)--(c) and coherent (d)--(f)
interaction of two pairs of counterpropagating beams. The initial
offset is $4x_0$ for in-phase beams propagating to the right in
(b) and (e), and $2x_0$ for the out-of-phase beams propagating to
the left in (c) and (f). Parameters and layout as in
Fig.~\ref{fig1}.}
\end{figure}
\begin{figure}
\includegraphics[width=75mm]{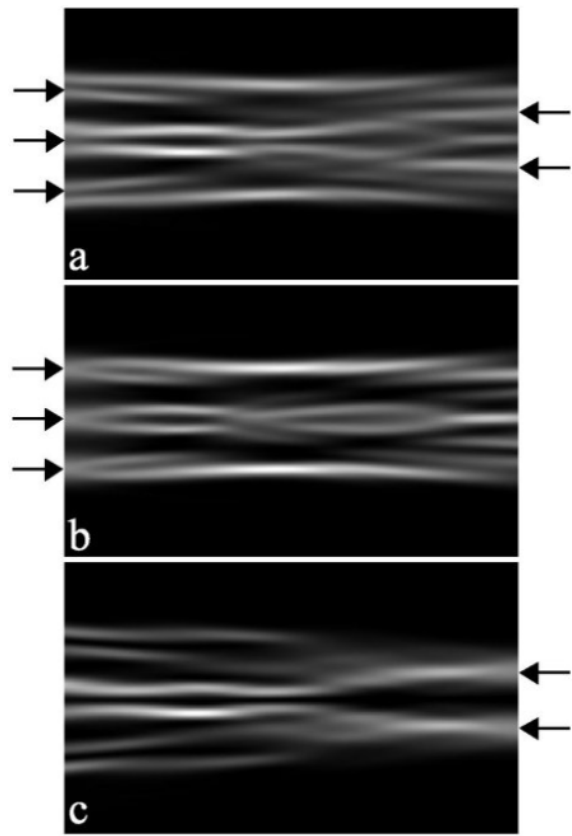}\vspace{-2mm}
\caption{\label{fig4} Unstable self-organized beam structure after
$116\tau_0$ temporal steps, at the moment of symmetry-breaking.
Three in-phase beams propagate to the right (b), two out-of-phase
beams to the left (c). The components have equal powers and the
coupling $\Gamma=10$. Other parameters as in Fig.~\ref{fig1}.}
\end{figure}

Selecting synchronous terms in the nonlinear paraxial wave
equation we obtain the propagation equations
\begin{subequations}
\begin{eqnarray}
i\partial_z F+\partial^2_x F=\Gamma\left[E_0F+E_1B/2\right]\ , \\
-i\partial_z B+\partial^2_x B=\Gamma\left[E_0B+E^{*}_1F/2\right],\
\end{eqnarray}
\end{subequations}
where the parameter $\Gamma=(k n_0 x_0)^2r_{eff}E_e$, and we use
the rescaling $x\rightarrow x/x_0$, $z\rightarrow z/L_D$,
$(F,B)\rightarrow(F,B)\exp(-i\Gamma z)$. Here $x_0$ is the typical
beam waist and $L_D=2kx_0^2$ is the diffraction length
\cite{comment}. Propagation equations are solved numerically,
concurrently with the temporal equations. The numerical procedure
consists of solving Eqs.~(\ref{temp}) for the components of the
space-charge field with the light fields obtained at every step as
a \emph{guided modes} of common induced waveguide. This is
achieved by an internal relaxation loop, i.e.\ nested within the
temporal loop, based on a beam-propagation method for the right-
and left-propagating components. Both loops are iterated until
convergence, which however is not necessarily reached in the
temporal loop. In that case a \emph{dynamical} state is obtained.

Head-on collision of the beams with initial soliton profiles,
after temporal relaxation to a steady-state, results in the
formation of a counterpropagating soliton (not shown), similar to
the one found in Ref.~\cite{cohen}. One can easily generalize this
approach, introducing higher-order counterpropagating solitons,
similar to the multihump vector solitons in co-propagating
geometry (see, e.g. Ref.~\cite{lena}). In Fig.~\ref{fig1} we
present a particular case of a dipole-mode counterpropagating
soliton. Dipole beam is launched from the right, and a
power-matched single beam from the left. Such a bimodal
counterpropagating soliton has been studied in Ref.~\cite{haelt}
and has been found to be stable in co-propagating geometry
\cite{lena}. The size of data windows in all figures is $10$ beam
diameters transversely by $2$ diffraction lengths longitudinally.

Shooting initial beams with arbitrary parameters generally leads
to $z$-dependent or non-stationary character of the beam
propagation. In some domain of the initial parameters, for example
with the relative angle of beam scattering $\theta$ close to $\pi$
and small initial transverse offset, our time-relaxation procedure
converges to the \emph{stationary in time} structures, which we
denote as steady-state \emph{self-trapped waveguides}. The
formation of a single bidirectional waveguide is shown in
Fig.~\ref{fig2}. Two coherent Gaussian beams are launched at
different lateral positions perpendicular to the crystal edges,
$\theta =\pi$. Both beams diffract initially, until the
space-charge field is developed in time to form the waveguide
induced by the total light intensity, Fig.~\ref{fig2} (a), and
this induced waveguide traps both beams, Fig.~\ref{fig2} (b) and
(c). When the initial separation is $4$ or more beam diameters,
the beams hardly feel the presence of each other, and focus into
individual solitons. For the separation of $2$ beam diameters, the
interaction is strong enough for the beams to form a joint
waveguiding structure, as it shown in Fig.~\ref{fig2}.
\begin{figure*}
\includegraphics[width=170mm]{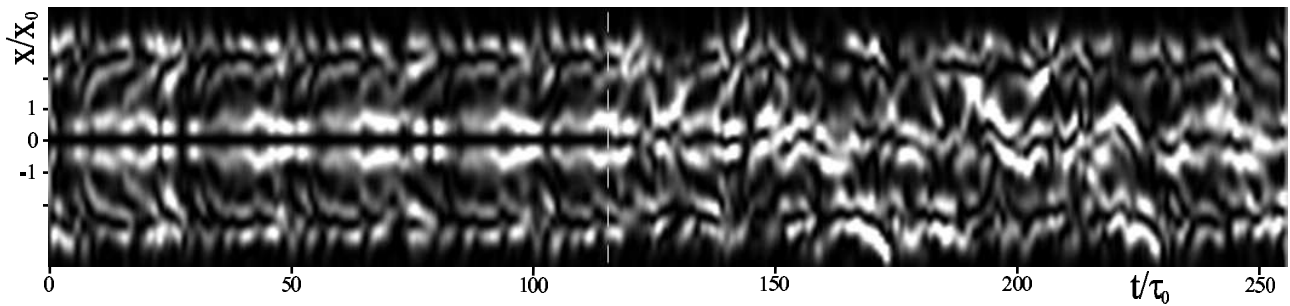}\vspace{-3mm}
\caption{\label{fig5} Temporal evolution of the output intensity
distribution of the two-lobe left-propagating beam at the left
face of the crystal. Dashed line at $t=116\tau_0$ shows the slice
corresponding to Fig.~\ref{fig4} (c), where the modulational
instability breaks the transverse symmetry.}
\end{figure*}

We would like to note here, that for a propagation distances
exceeding some threshold value, i.e.\ for larger crystal lengths,
we observe a longitudinal modulational instability developing in
time even for the initial beams corresponding to the exact
steady-state solitons. Modulational instability is a topic of
ongoing research and beyond the frame of present paper.

Having found different steady-state self-trapped structures, we
examine the difference between the coherent and incoherent
interaction of beams. Two steady-state solutions with the same
boundary conditions but for different degrees of mutual coherence
$\varepsilon$ are shown in Fig.~\ref{fig3}. Counterpropagating
beam components made of two pairs of beams are launched with a
lateral offset. The beams to the right are in-phase, and aim at
the center of the opposite crystal face. The beams to the left are
out-of-phase, and launched in parallel. Figs.~\ref{fig3} (a)--(c)
depict the incoherent interaction, $\varepsilon=0$. The beams
attract, focus and overlap tightly, but the ones to the right (b)
are still capable of building the intense spot in between the
other two. However, in the coherent case $\varepsilon=1$, shown in
Fig.~\ref{fig3} (d)--(f), beams focus and overlap less, and the
beams to the right (e) are expelled from the region between the
other beams. Also, the time scale of build-up dynamics is shorter
for coherent then for incoherent beams.

Of special interest are those self-trapped structures that
dynamically do not converge to a steady state. Such structures
represent novel time-dependent, as well as $z$-dependent
waveguides that can not be described by the usual steady-state
theory of spatial solitons. Whereas the $z$-dependence can be
ascribed to the general definition of longitudinal waveguide
modes, the time-dependence is a novel feature, caused by the slow
response of PR crystals. An example is depicted in
Fig.~\ref{fig4}, where a collision of three against two
power-matched coherent beams is presented. The initial
configuration is such that the three beams propagating to the
right interfere constructively (b), to overlap with the two
counterpropagating out-of-phase beams (c). These two beams,
propagating to the left, repel and overlap with the two
outside-lying opposite beams. During the time evolution of this
dynamical state we have observed several alternations of
transversely symmetrical structures, similar to the one shown in
Fig.~\ref{fig4}, and identified such behavior as a quasi-periodic
self-oscillation \cite{silberberg}, clearly seen in
Fig.~\ref{fig5} for $t<116\tau_0$. At that point the development
of \emph{transverse} symmetry-breaking instability is observed,
which results in irregular dynamics, shown in Fig.~\ref{fig5} for
$t>116\tau_0$.

In conclusion, we have developed a theory of self-trapped
bidirectional waveguides. In counterpropagating geometry, the
inclusion of time dependent effects was found to be crucial for
the formation of joint waveguiding structures. We demonstrated the
generation of a counterpropagating (1+1)D vector soliton
numerically and proposed a more general class of non-soliton
steady-state solutions. The level of temporal coherence of
interacting beams influences the mutual coupling due to the
formation of a refractive index grating. In addition to the
generation of steady-state induced waveguides, the dynamic
alternation of states following by transverse modulational
instability, as well as the onset of longitudinal modulational
instability were observed.\medskip

\begin{acknowledgments}\vspace*{-5mm}
MB, AS, and AD gratefully acknowledge support from the Alexander
von Humboldt Foundation for the stay and work at WWU Muenster.
Part of this work was supported by the Deutsche
Forschungsgemeinschaft.
\end{acknowledgments}


\begin{thebibliography}{}
\bibitem{comment} Propagation equations can be put in a universal
dimensionless form that contains no parameters or coupling
constants. All the parameters are then hidden in the scaling
quantities and the initial and boundary conditions. We prefer the
form with one driving coupling constant.
\end{thebibliography}
\end{document}